\documentclass{aa}
\usepackage{graphicx}

\begin{document}

\def\etal{et al. }
\def\araa{Ann.\ Rev.\ Astron.\ Ap.}
\def\aplet{Ap.\ Letters}
\def\aj{Astron.\ J.}
\def\apj{ApJ}
\def\apjl{ApJ\ Lett.}
\def\apjs{ApJ\ Suppl.}
\def\aas{A\&A Supp.}
\def\aa{A\&A}
\def\aal{A\&A Lett.}
\def\mnras{MNRAS}
\def\mnrasl{MNRAS Lett.}
\def\nature{Nature}
\def\apss{Ap\&SS}
\def\pasa{{\it Proc.\ Astr.\ Soc.\ Aust.}}
\def\pasp{{\it P.\ A.\ S.\ P.}}
\def\pasj{{\it PASJ}}
\def\pre{{\it Preprint}}
\def\aph{Astro-ph}
\def\sovlet{{\it Sov. Astron. Lett.}}
\def\adspr{{\it Adv. Space. Res.}}
\def\expas{{\it Experimental Astron.}}
\def\ssr{{\it Space Sci. Rev.}}
\def\ar{{\it Astronomy Reports}}
\def\inpress{in press}
\def\inprep{in prep.}
\def\submit{submitted}

\def\ap{$\approx$ }
\def\mjysr{MJy/sr }
\def\inu{{I_{\nu}}}
\def\inufit{I_{\nu fit}}
\def\fnu{{F_{\nu}}}
\def\bnu{{B_{\nu}}}
\def\msol{{M$_{\odot}$}}
\def\mic{$\mu$m}
\def\cm2{$cm^{-2}$}

\title{Inverse temperature dependence of the dust submillimeter spectral index}

\author{X. Dupac\inst{1,2}, J.-P. Bernard\inst{1},
  N. Boudet\inst{1}, M. Giard\inst{1}, J.-M. Lamarre\inst{4}, C. M\'eny\inst{1},
  F. Pajot\inst{3},
  I. Ristorcelli\inst{1}, G. Serra\inst{1}, B. Stepnik\inst{3,1}, J.-P. Torre\inst{5}}
\institute{Centre d'\'Etude Spatiale des Rayonnements, 9 av. du Colonel Roche, BP4346, F-31028 Toulouse cedex 4, France
\and
European Space Agency - ESTEC, Astrophysics Division, Keplerlaan 1, 2201 AZ Noordwijk, The Netherlands
\and
Institut d'Astrophysique Spatiale, Campus d'Orsay, b\^at. 121, 15 rue Cl\'emenceau, F-91405 Orsay cedex, France
\and
LERMA, Observatoire de Paris, 61 av. de l'Observatoire, F-75014 Paris, France
\and
Service d'A\'eronomie du CNRS, BP3, F-91371 Verri\`eres-le-Buisson cedex, France
}

\offprints{xdupac@rssd.esa.int}

\authorrunning{Dupac \etal}
\titlerunning{Temperature dependence of the dust spectral index}

\date{Received {} /Accepted {}}

\abstract{We present a compilation of PRONAOS-based results concerning the temperature dependence of the dust submillimeter spectral index, including data from Galactic cirrus, star-forming regions, dust associated to a young stellar object, and a spiral galaxy.
We observe large variations of the spectral index (from 0.8 to 2.4) in a wide range of temperatures (11 to 80 K).
These spectral index variations follow a hyperbolic-shaped function of the temperature, high spectral indices (1.6-2.4) being observed in cold regions (11-20 K) while low indices (0.8-1.6) are observed in warm regions (35-80 K).
Three distinct effects may play a role in this temperature dependence: one is that the grain sizes change in dense environments, another is that the chemical composition of the grains is not the same in different environments, a third one is that there is an intrinsic dependence of the dust spectral index on the temperature due to quantum processes.
This last effect is backed up by laboratory measurements and could be the dominant one.
}

\maketitle


\section{Introduction\label{intro}}

Characterizing dust emissivity properties is one of the major challenges of nowadays submillimeter astronomy.
In this spectral range, big grains at thermal equilibrium (see for instance \cite{desert90}) dominate the dust emission.
Their emission is characterized by a temperature and a spectral dependence of the emissivity which is usually simply modelled by a spectral index.

The temperature, density and opacity of a molecular cloud are key parameters which control the structure and evolution of the clumps, and therefore, star formation.
The spectral index ($\beta$) of a given dust grain population is directly linked to the internal physical mechanisms and the chemical nature of the grains.

It is generally admitted from Kramers-K\"onig relations that 1 is a lower limit for the spectral index (\cite{emerson88}).
A classical value is 2, following the investigations of Gezari \etal (1973) and the calculations of, for instance, Draine \& Lee (1984).
This value is particularly invoked for isotropic crystalline grains, and it is thought to be an upper limit (\cite{wooten72}, \cite{emerson88}).
For amorphous silicate or graphitic grains, a value of $\beta = 2$ is also favoured for different physical reasons (\cite{rowanrobinson86}, \cite{tielens87}, \cite{mennella95}).
However, it is not the case for amorphous carbon, which is thought to have a spectral index equal to 1 (\cite{koike80}), as well as aggregates of silicates and graphite in a porous structure (\cite{mathis89}).
Silicate grains embedded in ice mantles are thought to have spectral indices between 1.5 and 2 (\cite{miyake93}, \cite{preibisch93}, \cite{krugel94}, \cite{pollack94}).
In general, amorphous rather than crystalline structure and increase of the grain size (\cite{mannings94}) are the main reasons to predict submillimeter spectral indices lower than 2.
Spectral indices above 2 may exist, according to several laboratory measurements on grain analogs (\cite{koike95}, \cite{mennella95}, \cite{agladze96}, \cite{mennella98}).
Also, Agladze \etal (1996) and Mennella \etal (1998) showed an anticorrelation existing for some types of grains between the temperature and the emissivity spectral index.

Observations of the diffuse interstellar medium at large scales favour $\beta$ around 2 (see for example \cite{boulanger96}, \cite{dunne01}).
In the case of molecular clouds, spectral indices are usually found to be between 1.5 (\cite{walker90}) and 2 (e.g. \cite{wright92} in Orion).
However, low values (0.2-1.4) of the spectral index have been observed in circumstellar environments (\cite{weintraub89}, \cite{beckwith90}, \cite{knapp93}).
This can be attributed to grain growth in dense stellar envelopes.
Low indices have also been observed in molecular cloud cores (e.g. \cite{blake96}).
Spectral indices larger than 2 have also been observed in the millimeter range (\cite{schwartz82}, \cite{wilson95}, \cite{kuan96}), in particular on the whole sky at large scales by the WMAP satellite (\cite{bennett03}).

In this context, submillimeter multi-band imaging of the interstellar medium provides a lot of useful information about dust properties and interstellar medium structure, especially if the dust emission
parameters, namely both the temperature and the spectral index, can be properly derived on top of submillimeter intensities.

\section{Observations\label{obs}}

PRONAOS (PROgramme NAtional d'Observations Submillim\'etriques) is a French
balloon-borne submillimeter experiment (\cite{serra01}, \cite{ristorcelli98}, \cite{lamarre94}).
Four bolometers cooled at 0.3 K measure the submillimeter flux with sensitivity to low
brightness gradients of about 4 \mjysr in band 1 (200 \mic) and 0.8 \mjysr in band 4 (580 \mic).
The effective wavelengths are 200, 260, 360 and 580 \mic, and the angular resolutions are
2$'$ in bands 1 and 2, 2.5$'$ in band 3 and 3.5$'$ in band 4.
The data that we analyze here were obtained during the second flight of PRONAOS in September 1996, at Fort Sumner, New Mexico.
The data processing method, including deconvolution from chopped data, is described in Dupac \etal (2001), and
the calibration procedure is detailed in Pajot \etal (2003).
This experiment has observed various phases of the interstellar medium, from diffuse clouds in Polaris (\cite{bernard99}) and Taurus (\cite{stepnik03}) to massive star-forming regions in Orion (\cite{ristorcelli98} and \cite{dupac01}), Messier 17 (\cite{dupac02}), Cygnus B, and the dusty envelope surrounding the young massive star GH2O 092.67+03.07 in NCS (\cite{bernard03}).
The $\rho$ Ophiuchi low-mass star-forming region has also been observed, as well as the edge-on spiral galaxy NGC 891 (\cite{dupac03}).
In this letter, we analyze constraints that PRONAOS data establish on the dust spectral index and its relation to the temperature.
Ristorcelli \etal (1998) and Bernard \etal (1999) gave tentative evidences of a spectral index larger than 2, respectively in Orion and the Polaris flare.
Dupac \etal (2001) and Dupac \etal (2002) showed large variations of the spectral index, including values above 2, and the presence of an anticorrelation between the temperature and the spectral index in massive star-forming regions.
In this letter, we extend the analysis to other regions observed with PRONAOS.

\section{Analysis\label{ana}}

\subsection{Bases of the analysis}

For Orion (\cite{dupac01}), M17 (\cite{dupac02}) and Cygnus (\cite{boudet03}), we analyze the intensity maps using the following procedure.
With the PRONAOS and IRAS intensity maps, we produce temperature ($T$) and spectral index ($\beta$) maps following the processing described in Dupac \etal 2002 (see also \cite{dupac01}).
We fit a modified black body law to the spectra: $\inu = \epsilon_0 \; \bnu(\lambda,T) \; (\lambda/\lambda_0)^{-\beta}$, where $\inu$ is the spectral intensity (MJy/sr), $\epsilon_0$ is the emissivity at $\lambda_0$ of the observed dust column density, $\bnu$ is the Planck function, $T$ is the temperature and $\beta$ is the spectral index.
In most of the areas, we use either only PRONAOS data or PRONAOS + IRAS 100 \mic~data.
The reasons why we do not use IRAS data everywhere are that first, some very intense regions are saturated in the 100 \mic~IRAS band, second, many faint regions which are clearly detected with PRONAOS are very noisy with IRAS, or a large-scale warm component dominates in the IRAS bands.
In the non-saturated warmest areas ($>$ 70 K), we use IRAS 60 \mic~data too (see \cite{dupac02} for more details).

\begin{figure}[]
\includegraphics[clip=true,viewport=10 10 500 350,scale=.5]{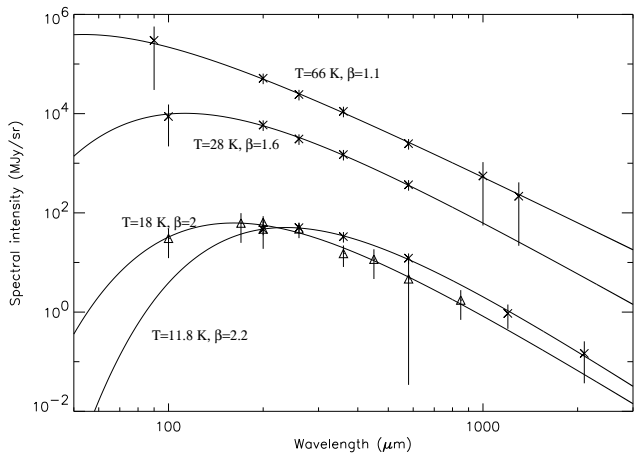}
\caption[]{Examples of spectra.
From top to bottom, the crosses represent data from the Orion Molecular Cloud - 1, M17 North, and Cloud 2 in Orion (see \cite{dupac01} for details).
Triangles represent data from the south-western region of the NGC 891 galaxy (see \cite{dupac03}).
The error bars are plotted within the 3 $\sigma$ confidence level.
The full lines represent the results of the fit by a single modified black body.
The parameters of each fit (temperature and spectral index) are also indicated.}
\label{spectra}
\end{figure}

In this work, we restrain the analysis to all fully independent (3.5$'$ side) pixels for which both relative errors on the temperature and the spectral index are less than 20\%.
This procedure is motivated notably by the fact that there exists a degeneracy between the temperature and the spectral index.
This effect is mainly due to the relatively large error bars on the spectral index in cold regions and on the temperature in warm regions (Rayleigh-Jeans limit).
Dupac \etal (2001) and Dupac \etal (2002) have shown by fitting simulated data that this artificial anticorrelation effect was small compared to the effect observed in the data.
For this work, we have also done other simulations, in a slightly different way from the previous ones: we consider temperature values in the same temperature range as the observations, but with a constant spectral index (1.5), we produce simulated data with a realistic noise level, and we fit them with the usual procedure.
With PRONAOS simulated data alone, we find that the relative residual error on the spectral index is 22\% rms.
With PRONAOS + IRAS 100 \mic~data, the same error is only 7\% rms on the spectral index and 9\% on the temperature.
The induced inverse correlation effect is very limited.
Thus, this noise effect is clearly not capable to explain the $T$-$\beta$ distribution found in the data.
Here we still reduce it by restraining the analysis to pixels for which errors on both the temperature and the spectral index are lower than 20\%.
Also, we disregard pixels for which at least one of the bands has a spectral intensity below 10 MJy/sr.
Moreover, all spectra were visually checked in order to exclude pixels which could be contaminated by noise features.
This provides us with a set of 31 pixels for M17, 19 for Orion and 60 for Cygnus.
For the cirrus clouds in Polaris and Taurus, we have, respectively, two and one data points available (\cite{bernard99} and \cite{stepnik03}).
For the NCS data point, we fit the PRONAOS fluxes as given in Bernard \etal (2003).
We also present five data points from $\rho$ Ophiuchi (\cite{ristorcelli03}) and three from the NGC 891 edge-on spiral galaxy (\cite{dupac03}).
We present typical examples of the obtained spectra in Fig. \ref{spectra}.

\subsection{Dependence on the temperature of the spectral index}

\begin{figure}[]
\includegraphics[scale=.5]{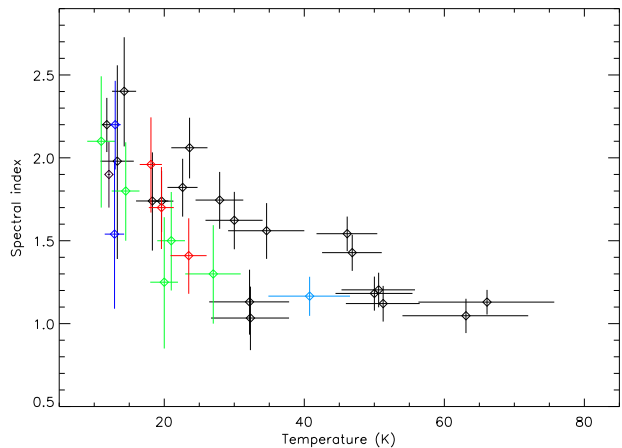}
\caption[]{Spectral index versus temperature, for fully independent pixels in Orion (black), $\rho$ Ophiuchi (green), Polaris (blue), Taurus (purple), NCS (light blue) and NGC 891 (red).
The error bars are plotted within the 68\% confidence interval.
For clarity, the M17 and Cygnus points are not presented in this figure, but they have similar error bars as the Orion points.
}
\label{compil}
\end{figure}

\begin{figure}[]
\includegraphics[scale=.5]{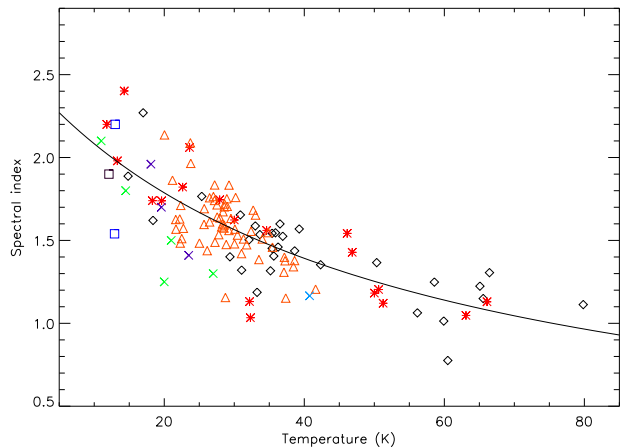}
\caption[]{Spectral index versus temperature, for fully independent pixels in Orion (red asterisks), M17 (black diamonds), Cygnus (orange triangles), $\rho$ Ophiuchi (green crosses), Polaris (blue squares), Taurus (purple square), NCS (light blue cross) and NGC 891 (purple crosses).
The full line is the result of the best hyperbolic fit: $\beta = {1 \over 0.4 + 0.008 T}$
}
\label{compil1}
\end{figure}

We present in Fig. \ref{compil} and \ref{compil1} temperature versus spectral index plots of the above-defined pixel set.
For clarity, we present all points in Fig. \ref{compil1} without the error bars, and a plot with the error bars in Fig. \ref{compil} but without the numerous points from Cygnus and M17.
These data points have similar error bars to the Orion data points.
The temperature in this data set ranges from 11 to 80 K, and
the spectral index also exhibits large variations from 0.8 to 2.4.
One can observe an anticorrelation on these plots between the temperature and the spectral index, in the sense that the cold regions have high spectral indices around 2, and warmer regions have spectral indices below 1.5.
In particular, no data points with $T >$ 35 K and $\beta >$ 1.6 can be found, nor points with $T <$ 20 K and  $\beta <$ 1.5.
All cold regions below 18 K have spectral indices above 1.8, except one point with a large error bar, which is well compatible with a value above 1.8.

This anticorrelation effect is present for all objects in which we observe a large range of temperatures, namely Orion, M17, Cygnus and $\rho$ Ophiuchi.
It is also remarkable that the few points from other regions are well compatible with this general anticorrelation trend.
Following the type of the observed object, the temperature distribution of the points is different, but the points seem to follow this same general trend of an inverse dependence of the spectral index on the temperature.
Hereafter, we therefore consider that this effect can be simply investigated with the different objects altogether.

Fitting the data under the assumption of a constant spectral index leads to $\chi^2$ / degrees of freedom (d.o.f. hence on) values of 1835/121 for $\beta$=2 and 377/121 for $\beta$=1.5.
These high values of the $\chi^2$/d.o.f. exclude the possibility of matching the data with an invariant spectral index.
The linear regression fit of $\beta$($T$), considering the temperature as an independent variable, gives a slope of $-$0.014 K$^{-1}$, and $\beta$ = 2.01 at $T$ = 0.
Then the linear correlation coefficient is $-$0.79, and the $\chi^2$/d.o.f. is 137/120 (=1.14).

Since the points seem quite well distributed along a hyperbolic-shape line, we perform a hyperbolic fit of the data using: $\beta = {1 \over \delta + \omega T}$, where $\delta$ and $\omega$ are free parameters.
Considering that the temperature is the independent variable, the best fit gives $\delta$=0.40 $\pm$ 0.02 and $\omega$=0.0079 $\pm$ 0.0005 K$^{-1}$, with $\chi^2$/d.o.f. = 120/120.
The temperature dependence of the emissivity spectral index is thus very well fitted with an hyperbolic approximating function.
Note that since the number of free parameters is the same as in the linear regression, the lower $\chi^2$ is significant, so that we can consider the hyperbolic fit to be better.

If now we consider that the spectral index is the independent variable, fitting the $T$($\beta$) function by the linear regression gives a large $\chi^2$/d.o.f. value of 1099/120.
In this case, the hyperbolic fit gives $\chi^2$/d.o.f. = 1123/120.
These bad fits are mainly due to the large dispersion of the spectral indices that we observe at low temperatures.

\section{Discussion}

In order to interpret this inverse temperature dependence of the spectral index, several hypotheses can be investigated.
Mixtures of dust components with different temperatures, but having the same spectral index, can be put forward to explain apparent low indices.
However, this effect was shown to be small compared to the amplitude of the observed spectral index variations (\cite{dupac02}).

As stated in the introduction, low indices have been observed in active environments such as circumstellar disks and warm molecular cloud cores.
PRONAOS observations of Orion, M17, Cygnus B and $\rho$ Ophiuchi
show such low spectral indices around 1.
This could be due to the increase of the grain size in dense environments, as predicted by Miyake \& Nakagawa (1993), whose models with different grain size distributions show that the maximum grain radius adopted is crucial for the long-wavelength emissivity slope.
This is also the reason invoked by Goldsmith \etal (1997) to explain their results, which showed an inverse dependence of the spectral index on the optical depth.
However, we do not find such correlation in our data.
Moreover, the range of variations that they find for the spectral index (from 1.6 to 3.2) is significantly different from ours (from 1 to 2.4), though with a similar variation amplitude.
Also, these grain growth effects are invoked for very peculiar dense environments around stars rather than for large scale observations such as the PRONAOS ones.
Therefore, it is difficult to consider that size distribution effects are dominant in the $T$-$\beta$ anticorrelation that we observe, though they may play a role.

Another slightly different explanation could be that one finds different chemical composition or physical state of the grains following the physical conditions of the medium, in particular following the temperature.
Cold environments could harbor fluffy silicate grains including ice compounds, having spectral indices of 2 (e.g. \cite{krugel94}), or simply silicate or graphitic particles (e.g. \cite{draine84}), or even olivine core or fused quartz grains with ice mantles (\cite{aannestad75}, $\beta \approx$ 3) to explain high indices above 2.
Warm regions could harbor aggregates of silicates, porous graphite or amorphous carbon and therefore have a spectral index around 1 (\cite{mathis89}).
Nevertheless, more observations are needed to constrain the chemical structure of the large grains in various media and confirm this quite speculative hypothesis.

Finally, an explanation may be searched for in a temperature dependence of the intrinsic optical properties of the grains (see \cite{henning97} for a review of the different physical processes which could be involved).
This can be linked to laboratory results (\cite{agladze96} and \cite{mennella98}) which showed a T-$\beta$ anticorrelation on interstellar grain analogs.
Mennella \etal (1998) measured the absorption coefficient of crystalline and amorphous grain analogs between 20 \mic~and 2 mm
wavelength, in the temperature range 24-295 K. They derived an
anticorrelation between the temperature and the spectral index, and attributed it to two-phonon
difference processes (\cite{sparks82}).
Agladze \etal (1996) measured absorption spectra of silicates between 0.7 and 2.9 mm wavelength and deduced an anticorrelation
between the spectral index and the temperature in the temperature range 10-25 K, for two different silicate precursors.
They attributed the underlying absorption process to a resonant tunneling effect between ground states of two-level systems (\cite{phillips72}).
The temperature dependence of the spectral index which they show is in a range of temperatures and spectral indices close to our measurements: they measure spectral indices between 1.8 and 2.6 for temperatures ranging from 10 to 25 K, while we measure spectral indices roughly between 1.4 and 2.4 in the same temperature domain.
However, the wavelength range is not the same, and there is no evidence that this effect is simply transposable from the millimeter range to the PRONAOS submillimeter range, since experimental results suggest a cut-off around 500 \mic~for this phenomenon (\cite{fitzgerald00}).
Furthermore, at submillimeter wavelengths and higher temperatures,
thermally-activated relaxation processes and temperature-dependent absorption associated to transitions to excited levels of two-level systems (\cite{fitzgerald00}, \cite{hutt89}) might dominate.
These issues will be treated in a forthcoming paper (M\'eny et al., in prep.)
This possible discrepancy between the submillimeter and the millimeter spectral indices could explain the small difference observed between the WMAP (millimeter) spectral index of 2.2 and the PRONAOS one, which is around 2 at DIRBE (rather low) large-scale temperatures.

\section{Conclusion}

We have presented a compilation of PRONAOS-based results concerning the temperatures and spectral indices measured in a variety of regions of the interstellar medium.
The temperatures observed range from 11 to 80 K, which provides a large field of investigation for measuring variations of the dust emissivity.
We give strong evidence of large variations of the submillimeter spectral index between 0.8 and 2.4, as well as of the existence of an inverse dependence of the emissivity spectral index on the temperature.
This dependence is well fitted by the function $\beta = {1 \over 0.4 + 0.008 T}$.
Several interpretations are possible for this effect: one is that the grain sizes change in dense environments, another is that the chemical composition of the grains is not the same in different environments and that this correlates to the temperature, a third one is that there is an intrinsic dependence of the spectral index on the temperature, due to quantum processes such as two-level tunneling effects.
Additional modeling, as well as additional laboratory measurements and astrophysical observations, are required in order to discriminate between these different interpretations.
However, the most likely explanation could be that there is an intrinsic dependence of the emissivity on the temperature.
A new dust semi-empirical model is needed to take into account these variations of the spectral index.

\section{Acknowledgements}
We thank very much H. A. Sabat for fruitful discussions on statistics and B. Parise for her careful reading of the manuscript.
We are indebted to the French space agency Centre National d'\'Etudes Spatiales
(CNES), which supported the PRONAOS project.
We are very grateful to the
PRONAOS technical teams at CNRS and CNES, and to the NASA-NSBF balloon-launching facilities group of Fort Sumner (New Mexico).

\end{document}